\documentclass[12pt,a4paper,titlepage,oneside]{article}
\usepackage{amsmath}
\usepackage{graphicx}%

\begin{document}

\begin{center}
{\bf \Large The quaternionic commutator bracket and its implications}
\end{center}

{\bf A. I. Arbab$^{1}$} and {\bf M. Al-Ajmi$^{2}$} \\

$^1$Department of Physics, Faculty of Science, University of
  Khartoum, P.O. Box 321, Khartoum 11115, Sudan \\
$^2$Department of Physics, College of Science, Sultan Qaboos
  University , P.O. Box 36, Alkhod 123, Oman \\

{\bf email:} aiarbab@uofk.edu, mudhahir@squ.edu.om \\

\section*{Abstract}
A quaternionic commutator bracket for position and momentum shows that
the quaternionic wave function, \emph{viz.}
$\widetilde{\psi}=(\frac{i}{c}\,\psi_0\,,\vec{\psi})$, represents a
state of a particle with orbital angular momentum, $L=3\,\hbar$,
resulting from the internal structure of the particle. This angular
momentum can be attributed to spin of the particle. The vector
$\vec{\psi}$, points along the direction of $\vec{L}$. When a charged
particle is placed in an electromagnetic fields the interaction energy
reveals that the magnetic moments interact with the electric and
magnetic fields giving rise to terms similar to Aharonov-Bohm and
Aharonov-Casher effects.

\section{Introduction}

In quantum mechanics particles are described by relativistic or
non-relativistic wave equations. Each equation associates a spin state
of the particle to its wave equation. For instance, Schrodinger
equation applies to the spin-0 particles in the non-relativistic
domain, while the appropriate relativistic equation for spin-0
particles is the Klein-Gordon equation.  The relativistic spin-1/2
particles are governed by the Dirac equation. Recall that this
equation reduces, in the non-relativistic limit, to the Schrodinger -
Pauli equation \cite{mess, drell}. Particles with spin -1 are
described by Maxwell equations. Charged particles interact with the
electromagnetic field. The effect of the interaction of these
particles is obtained via the \emph{minimal coupling} ansatz, where
the momentum of the free particle ($p_\mu$) is replaced by $p_\mu+q
A_\mu$, where $A_\mu$ is the photon field. With this prescription,
Dirac equation as well as the Schrodinger - Pauli equation predict the
existence of spin for the electron through the interaction of its
magnetic moment with magnetic field present. Thus, the spin angular
momentum is deemed to be a quantum phenomenon having no classical
analogue. Hence, the spin is an intrinsic property of a quantum
particle. Therefore, the spin is not a result of the rotation of a
point quantum particle (like the electron). It could be associated
with the space in which the particle is defined. In our new
formulation the particle is defined by the scalar and vector
wavefunctions. The spin angular momentum can be defined in terms of
these vector and scalar wavefunctions.

In a recent paper, we have unified the above three quantum wave
equations in a single equation. We call this equation a unified
quantum wave equation \cite{uq}.  In the present work, we would like
to investigate the nature of spin relying on our unified quantum wave
equation, and using the quaternionic commutator bracket between
position and momentum, \emph{viz.},
$\left[x_i\,,p_j\right]=i\hbar\,\delta_{ij}$ \cite{mess}. This bracket
is a fundamental cornerstone in formulating quantum mechanics.

Such a generalization led to interesting physical results pertaining
to the nature of the wave equation describing quantum particles. We
have found that an intrinsic angular momentum, related to its internal
nature a quantum particle, is associated with the quaternionic wave
function $\widetilde{\psi}$ that we recently found. It is generally
understood that spin is not due to the point-particle rotation. This
is true if we treat a particle as a point particle. But, we have
recently shown that the wavy nature of a quantum particle due to its
wavepacket nature undergoes an internal rotation \cite{uq}. A
wavepacket consists of two waves moving in opposite directions with
speed of light. If a quantum particle is thought of a wavepacket or an
extended object, then an internal rotation is plausible.

Our current investigations revealed that the scalar ($\psi_0$) and
vector ($\vec{\psi}$) wavefunctions describe a particle with internal
orbital angular momentum, $L=\hbar$\,; and that the vector
$\vec{\psi}$ is directed a long the direction of $\vec{L}$. This
internal angular momentum can be attributed to some kind of spin.
These may define the helicity states of the quantum particle. Such
states can be compared with spinor representation of Dirac equation.
When a charged particle is placed in an electromagnetic field, the
magnetic moments ineteract with both electric and magnetic fields
giving rise to Aharonov-Bohm and Aharonov-Casher effects.

\section{The fundamental commutator bracket}
The position and momentum commutation relation in quaternionic now reads
\begin{equation}
\left[\widetilde{X}\,, \widetilde{P}\right]\widetilde{\psi}=i\hbar\,\widetilde{\psi}\,,
\end{equation}
where
\begin{equation}
\widetilde{X}=\left(ict\,, \vec{r}\right)\,,\,\,\, \widetilde{P}=\left(i\frac{E}{c}\,, \vec{p}\right),\,\,\, \widetilde{\psi}=\left(\frac{i}{c}\,\psi_0\,, \vec{\psi}\right)\,.
\end{equation}
The multiplication of two quaternions, $\widetilde{A}=\left(a_0\,, \vec{a}\right),\,\widetilde{B}=\left(b_0\,, \vec{b}\right)$ is given by
\begin{equation}
\widetilde{A}\,\widetilde{B}=\left(a_0b_0-\vec{a}\cdot\vec{b}\,, a_0\vec{b}+\vec{a}b_0+\vec{a}\times\vec{b}\right)\,.
\end{equation}
Using eq.(3) and the fact that in quantum mechanics, $\vec{p}=-i\hbar\vec{\nabla}$ and $E=i \hbar\frac{\partial}{\partial t}$, and eq.(1) to get
\begin{equation}
\vec{L}\cdot\vec{\psi}=\frac{3\hbar}{c}\,\psi_0\,,
\end{equation}
\begin{equation}
\vec{L}\,\psi_0=3\hbar\,c\,\vec{\psi}\,,
\end{equation}
and
\begin{equation}
 \vec{L}\times\vec{\psi}=0\,.
\end{equation}
Equation (5) states $\vec{L}$ transform the scalar wavefunction
$\psi_0$ into the vector wavefunction $\vec{\psi}$\,. Notice however
that in quantum mechanics, $\vec{L}\times\vec{L}=i\hbar\vec{L}$. But
according to eq.(5) and (6), $\vec{L}\times\vec{L}=0$. Thus, the
physical meaning of $\psi$ has now become clear. We have recently
developed the quaternionic quantum mechanics but the physical meaning
of the quaternion wavefunction remained unsorted \cite{qqm, uq, dd}.

Now take the dot product of eq.(5) with $\vec{L}$ and use eq.(4) to get
\begin{equation}
L^2\psi_0=9\,\hbar^2\,\psi_0\,.
\end{equation}
Similarly, multiplying eq.(4) by $\vec{L}$ (from right) and using
eq.(5) and, the vector identity
$\vec{A}\times(\vec{A}\times{B})=\vec{A}(\vec{A}\cdot{B})-A^2\vec{B}$,
we obtain
\begin{equation}
L^2\vec{\psi}=9\,\hbar^2\,\vec{\psi}\,.
\end{equation}
Hence, the quaternion components, $\psi_0$ and $\vec{\psi}$ represent
a state of a particle with a total orbital angular momentum of
\begin{equation}
L=3\,\hbar\,.
\end{equation}
This is a quite interesting result. It seems that this angular
momentum arises from an internal degree of freedom. It may result from
a rotation of some internal structure of the particle. In Dirac's
theory the spin of the electron doesn't emerge from the equation to be
1/2, but is deduced to be 1/2 from the way the electron interacts with
the photon field.

We have shown recently that the unified quantum wave equations are \cite{dd}
\begin{equation}
\vec{\nabla}\cdot\vec{\psi}-\frac{1}{c^2}\frac{\partial \psi_0}{\partial t}-\frac{m_0}{\hbar}\,\psi_0=0\,,
\end{equation}
\begin{equation}
\vec{\nabla}\psi_0-\frac{\partial \vec{\psi}}{\partial t}-\frac{m_0c^2}{\hbar }\,\vec{\psi}=0\,,
\end{equation}
and
\begin{equation}
\vec{\nabla}\times\vec{\psi}=0\,.
\end{equation}
In the ordinary quantum mechanics, a particle is described by a scalar
or spinor, however, a particle is now described by a scalar and a
vector. In all a particle is described by a four component function.
In electromagnetism, the electromagnetic fields $\vec{E}$ and
$\vec{B}$ are vector field. But at the fundamental level these two
field are represented by a scalar field $\varphi$ and a vector field
$\vec{A}$, respectively. In this manner, at the fundamental level, a
particle should be described by some similar fields, which are here
$\psi_0$ and $\vec{\psi}$. This makes the analogy between the field
and particle representations symmetric. The electromagnetic wave is
transverse, i.e., $\vec{E} \bot \vec{B}\bot \vec{k}$, while a particle
wave is longitudinal. In our formulation, this feature is very clear.
Equation (6) states that the spin direction is along the direction of
the field $\vec{\psi}$. As for photons, which are described by their
polarization a quantum particle should have some similar ansatz, where
the spin is directly associated with the particle wavefunction. Thus,
the wavefunction incorporates the spin states. In Dirac theory the
spin is deduced from the interaction of the electron with the photon
magnetic field. In Schrodinger-Pauli theory the spin of the electron
is also deduced from their equation in the way the electron spin is
coupled to the photon magnetic field. In ordinary quantum mechanics,
the spin of the particle does not emerge from its wavefunction, but
eqs.(7) and (8) show that it does in our present formulation.

And if $\psi_0=-\vec{v}\cdot\vec{\psi}$, i.e., when $\vec{\psi}$ is
projected along the direction of motion, then \cite{an}
\begin{equation}
  i\,\hbar\frac{d\psi_0}{dt}=m_0c^2\psi_0\,,\qquad i\,\hbar\frac{d\vec{\psi}}{dt}=m_0c^2\vec{\psi}\,.
\end{equation}
Now differentiate eq.(5) with respect to time and use eq.(13) to obtain
\begin{equation}
\frac{d\vec{L}}{dt}=0\,.
\end{equation}
this implies that the orbital angular momentum is a constant of
motion. It is thus a conserved quantity.  Moreover, if we take the dot
product of eq.(11) with $\vec{L}$ and use eqs.(4), (5) and (10) then
\begin{equation}
\frac{\partial \vec{L}}{\partial t}\cdot\vec{\psi}=0\,.
\end{equation}
and if we use the fact that
\begin{equation}
\frac{d\vec{L}}{dt}=\frac{\partial \vec{L}}{\partial t}+\vec{v}\cdot\vec{\nabla}\vec{L}\,,
\end{equation}
 then eq.(15) implies that
\begin{equation}
\frac{d\vec{L}}{dt}\cdot\vec{\psi}=0\,.
\end{equation}
Thus, either $\vec{L}$ is conserved or the external torque,
$\vec{\tau}=\frac{d\vec{L}}{dt}$, is perpendicular to $\vec{\psi}$.
Thus, the internal rotation of the particle is due to the
self-interaction (internal constituents) of the particle.

Now apply the condition, $\psi_0=-\vec{v}\cdot\vec{\psi}$, in eqs.(4),
(5) and (6), and using eq.(9), yield
\begin{equation}
\hat{L}\cdot\vec{v}=-c\,,\qquad \hat{L}=\frac{\vec{L}}{L}\,.
\end{equation}
This shows that $\vec{L}$ is along the opposite direction of motion
and that the particle moves with speed of light.

Now let us choose
\begin{equation}
\vec{\psi}=\frac{1}{c}\,\vec{\sigma}\,\psi_0\,,
\end{equation}
where $\sigma$ are the Pauli matrices. In this case, $\vec{\psi}$
would represent a spin vector wave.  Apply eq.(18) in eqs.(5) and (6)
to obtain
\begin{equation}
\vec{L}=3\,\hbar\,\vec{\sigma}\,.
\end{equation}
If substitute eq.(19) in the condition
$\psi_0=-\vec{v}\cdot\vec{\psi}$, we will obtain
\begin{equation}
\vec{\sigma}\cdot\frac{\vec{v}}{c}=-1\,.
\end{equation}
This shows that the spin component is antiparallel to the direction of
motion. This case agrees with that in eq.(18). Hence, the particle is
left-handed!  It seems that there is some internal degree of freedom
(angular momentum) associated with quaternionic particles. Or
alternatively, that the quaternionic space has some twisting
properties.
\section{Interaction energy}
Let us now consider the interaction energy ($U$) of the magnetic
dipoles in the presence of electromagnetic fields. This energy is due
to spin ($\vec{S}$) and orbital angular momentum ($\vec{L}$) and the
corresponding moments associated with them, $\vec{\mu}_s$ and
$\vec{\mu}_\ell$. The quaternionic form of the total angular momentum
is defined as
\begin{equation}
\tilde{J}=(0\,,\vec{L}-i\,\vec{S})\,.
\end{equation}
The corresponding interaction energy can be written as \cite{magnetic}
\begin{equation}
\tilde{U}=(iU\,,\vec{\tau})=-\tilde{\mu}\tilde{F}\,,
\end{equation}
where
\begin{equation}
  \tilde{\mu}=(0\,, \vec{\mu}_\ell-i\,\vec{\mu}_s)\,,\qquad \tilde{F}=(0\,, \frac{\vec{E}}{c}+i\,\vec{B})\,.
\end{equation}
When the dipole magnetic moment is placed in an external magnetic
field, it experiences a torque ($\vec{\tau}$). The torque tends to
align the dipole with the field. But when the interaction energy is
constant, the dipole moment precesses with the magnetic field.
Substituting eq.(24) in eq.(23) and equating the real and imaginary
parts of the resulting equations yield
\begin{equation}
  U=-\vec{\mu}_\ell\cdot\frac{\vec{E}}{c}-\vec{\mu}_s\cdot\vec{B}\,, \qquad \vec{\mu}_\ell\cdot\vec{B}=\vec{\mu}_s\cdot\frac{\vec{E}}{c}\,,\qquad \vec{\tau}=\vec{\mu}_s\times\vec{B}+\vec{\mu}_\ell\times\frac{\vec{E}}{c}\,,\qquad \vec{\mu}_\ell\times\vec{B}=\vec{\mu}_s\times\frac{\vec{E}}{c}\,.
\end{equation}
Equation (25) encompasses all possible terms that may arise due to the
presence of $\vec{\mu}_s$ and $\vec{\mu}_\ell$. The interaction of a
magnetic moment with electric field has not been considered widely by
physicists \cite{electric, ahar}, and hence this term is deemed to
vanish. This term violates parity and time reversal invariance but
respects rotational invariance \cite{electric}. Owing to duality
between electric and magnetic fields, we trust such terms should be
present. It is shown by Aharonov-Casher that a phase shift occurs for
a neutral particle with a nonzero magnetic dipole moment moving in an
electric field \cite{ahar}. This is given by
\begin{equation}
\Delta\varphi_{sE}=\oint(\vec{\mu}_s\times\vec{E})\cdot d\vec{r}\,.
\end{equation}
Using eq.(25) this is transformed into
\begin{equation}
\Delta\varphi_{\ell B}=\oint (c\,\vec{\mu}_\ell\times\vec{B})\cdot d\vec{r}\,.
\end{equation}
This states that the effect of a spin magnetic moment in an electric
field is equivalent to an orbital magnetic moment in a magnetic field.
Aharonov-Bohm also showed that in the absence of electric field in the
region, the phase shift of the particle wavefunction is given by
\begin{equation}
\Delta\varphi=-qV\Delta t/\hbar\,,
\end{equation}
where $V$ is the electric potential and $\Delta t$ is the time spent
in the electric potential \cite{aharnov}. In magnetic Aharonov the
phase shift is given by
\begin{equation}
\Delta\varphi=q\phi_B/\hbar\,,
\end{equation}
where $\phi_B$ is the magnetic flux enclosed by the solenoid.
The magnetic flux can be written as
\begin{equation}
\phi_B=\int\vec{B}\cdot d\vec{A}\,,
\end{equation}
but if
\begin{equation}
 \frac{d\vec{A}}{dt}=\frac{\vec{L}}{2m}=const.,\qquad \vec{A}=\frac{\vec{L}}{2m}\,\Delta t\,,
\end{equation}
reflecting conservation of the orbital angular momentum in the closed
loop, then applying eqs.(30) and (31) in eq.(29) yields
\begin{equation}
\Delta\varphi=\frac{q\,t}{2\hbar m}\vec{L}\cdot\vec{B}=\frac{1}{\hbar}\,\vec{\mu}_\ell\cdot\vec{B}\,\Delta t\,.
\end{equation}
Comparing eqs.(28) and (32) reveals that
\begin{equation}
\frac{d\varphi}{dt}=\frac{qV}{\hbar}\,.
\end{equation}
Hence, in the absence of electric field, the charge behaves as a
magnetic dipole interacting with a magnetic field. We thus may
attribute the effect above to the spin and orbital angular momentum
that the charge carries. It is of importance to remark that the rate
of change phase in the Josephson junction, that is mediated by Cooper
pairs, amounts to double the value in eq.(33) \cite{jose}.

\end{document}